Research Paper

# Fine Tuning Hydrophobicity of Counter-Anions to Tailor Pore Size in Porous All-Poly(ionic liquid) Membranes

*Zhiping Jiang[a], Yu-ping Liu[b], Yue Shao[a], Peng Zhao[a], Jiayin Yuan[c], Hong Wang[a]\**

―――――

\* Correspondence to: H Wang, College of chemistry, Institute of Polymer Chemistry, Key Laboratory of Functional Polymer Materials (Ministry of Education); Nankai University, Tianjin, 300071, P. R. China

Email: hongwang1104@nankai.edu.cn

[a] College of chemistry, Institute of Polymer Chemistry, Key Laboratory of Functional Polymer Materials (Ministry of Education); Nankai University, Tianjin, 300071, P. R. China

[b] College of Chemistry, Key Laboratory of Biosensing and Molecular Recognition (Tianjin), and Key Laboratory of Advanced Energy Materials Chemistry (Ministry of Education); Nankai University, Tianjin 300071, China;

[c] Department of Materials and Environmental Chemistry, Stockholm University, Stockholm, 10691, Sweden;

―――――

**Abstract:** Charged porous polymer membranes (CPMs) emerging as a multifunctional platform for diverse applications in chemistry, materials science, and biomedicine have been attracting widespread attention. Fabrication of CPMs in a controllable manner is of particular significance for optimizing their function and maximizing practical values. Herein, we report the fabrication of CPMs exclusively from poly(ionic liquid)s (PILs), and their pore size and wettability were precisely tailored by rational choice of the counter-anions. Specifically, stepwise subtle increase in hydrophobicity of the counter-anions by extending the length of fluorinated alkyl substituents, *i.e.* from bis(trifluoromethane sulfonyl)imide (Tf$_2$N) to bis(pentafluoroethane sulfonyl)imide (Pf$_2$N) and bis(heptafluoropropane sulfonyl)imide (Hf$_2$N), decreases the average pore size gradually from 1546 nm to 157 nm and 77 nm, respectively. Meanwhile, their corresponding water contact angles increased from 90º to 102º and 120º. The exquisite control over the porous architectures and surface wettability of CPMs by systematic variation of the anion's hydrophobicity provides a solid proof of the impact of the PIL anions on CPMs' structure.

## INTRODUCTION

There is currently much interest in devising physical and chemical methods for fabricating charged porous polymer membranes (CPMs) because the synergy of pore confinement,[1] charge character, and flexible chemical structure design of CPMs endows them with versatile applications.[2] Examples include but are not limited to, separator in fuel cell,[3] device fabrication,[4] separation,[5] controlled release,[6] catalyst supports,[7] bio-interfacing,[8] and sensors.[9] However, the charged nature of CPMs inherently makes them difficult to obtain by well-established methods in current industry. To this end, much effort has been devoted towards the exploration of new fabrication methods. So far, there are generally three effective strategies to fabricate CPMs: (i) the self-assembly and dewetting of block copolymers or their blends on a substrate to obtain CPMs;[10] (ii) electrostatic layer-by-layer assembly of specific building blocks of opposite charge under carefully designed conditions to make CPMs.[11] Unfortunately, comparatively large time and labor demand in these two methods and the difficulties to obtain freestanding interconnected CPMs hinders their real-life usage. (iii) Electrostatic complexation strategy was lately applied to fabricate free-standing CPMs by assembly of hydrophobic polycations and hydrophilic polyanions.[12] This freshly emerging synthetic route is powerful in pore size control while the surface characteristics are ignored. In fact, the surface characteristics (e.g., wettability, charge, and chemistry) are equally important to the pore systems in bulk CPMs for applications.

Poly(ionic liquid)s (PILs) are a class of ionic polymers using ionic-liquid-based moieties to build up polymer chains. The freedom in combining cations and anions in ionic liquid chemistry plus the macromolecular architectures opens up endless possibilities in a much wider window of their physicochemical properties in comparison to conventional polyelectrolytes.

In this study, we fabricated for the first time all-PIL-based CPMs, while previous PIL-based CPMs were produced exclusively from a mixture of PIL and poly(acrylic acid) (or other neutral acid compounds). The all-PIL method allowed for precise tuning of pore sizes and surface water wettability of CPMs in term of hydrophobicity of counter-anions, which was gently adjusted by the length of fluoridated alkyl substituents in the counter-anions. It is noteworthy, such method can be scaled up, and complement the current CPM fabrication technologies.

## EXPERIMENTAL

Experimental details including the synthesis of the polymers and the fabrication and analysis of the membranes are included in the Supporting Information.

## RESULTS AND DISCUSSION

As shown in **Figure 1**, we first synthesized a COOH-free PIL poly(1-cyanomethyl-3-vinylimidazolium bromide) (termed **PCMVImBr**), and a COOH-bearing PIL poly(1-carboxymethyl-3-vinylimidazolium bromide) (termed **PCAVImBr**). In the initial step, 1-vinylimidazole went quaternization reaction with 2-bromoacetonitrile and bromoacetic acid in acetone at room temperature to form 1-vinyl-3-cyanomethylimidazolium bromide (VCMImBr) and 1-vinyl-3-carboxymethyl imidazolium bromide (VCAImBr) monomers, respectively. For structural characterization and synthetic details, please refer to supporting information (Figure S1-S10). These polymers, PCMVImBr and PCMVImBr were synthesized by free radical polymerization of VCMImBr and VCAImBr in DMSO solution, respectively.

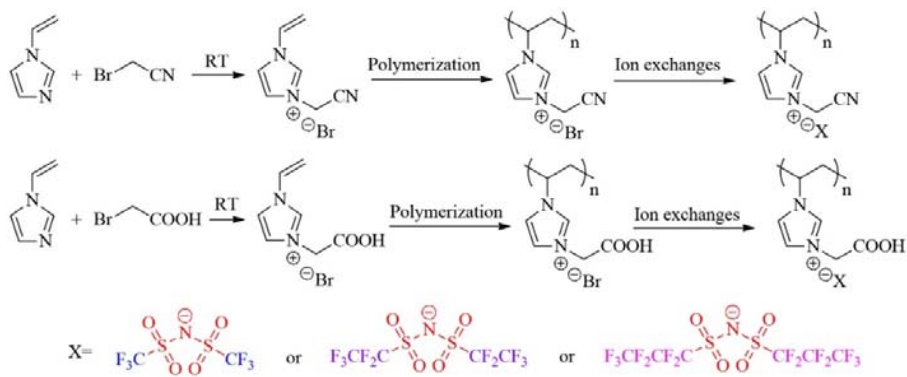

**Figure 1**. Synthetic routes towards PCMVImX and PCAVImX, where X represents $Tf_2N$, $Pf_2N$ and $Hf_2N$, respectively.

**Figure 2** shows the proton nuclear magnetic resonance ($^1$H-NMR) spectra of PCMVImBr and PCAVImBr in DMSO-$d_6$. In comparison to $^1$H-NMR spectra of their monomers (**Figure**

**S1**, **Figure S3**), all vinyl proton signals vanished and a broad signal between 1.5 and 3.2 ppm arises due to the newly formed polymer backbone. The rest proton signals became broader and can be assigned to individual protons on the polymer skeleton, indicative of successful polymerization. PCMVImBr and PCAVImBr were further analyzed by static light scattering (SLS) measurements, and their absolute weight-averaged molecular weights were determined to be $2.2 \times 10^5$ (polydispersity, 3.68) and $1.09 \times 10^5$ g/mol (polydispersity, 3.02), respectively. Next, PCMVImX and PCAVImX (x = $Tf_2N$, $Pf_2N$ and $Hf_2N$, respectively) were easily prepared by anion exchange reactions of PCMVImBr and PCAVImBr with the corresponding inorganic salts in an aqueous medium, respectively.

Thermal properties of PCMVImX and PCAVImX were analyzed by differential scanning calorimetry (DSC) and thermal gravimetric analysis (TGA). As shown in **Figure 2b**, the glass transition temperatures ($T_g$s) of PCMVImTf$_2$N, PCMVImPf$_2$N, PCMVImHf$_2$N are 104 °C, 112 °C and 109 °C, respectively. The $T_g$s of PCAVImTf$_2$N, PCAVImPf$_2$N, PCAVImHf$_2$N are 153 °C, 160 °C and 154 °C, respectively. The $T_g$s of PCMVImX are generally lower than that of PCAVImX because of strong hydrogen bonding enabled by the carboxyl-group in PCAVImX. In each polymer group, the $T_g$s are close to each other (within 10 °C) due to high structural similarity of $Tf_2N$, $Pf_2N$ and $Hf_2N$ anions, thus we could vary hydrophobicity of counter-anions without affecting much of the cation-anion interaction. This effect is ideal when investigating the relationship between hydrophobicity of counter-anions and pore size of CPMs with minimum structural change. The thermogravimetric analysis shows that all PIL-X studied here possess a moderate thermal stability up to 280 °C (**Figure S11**).

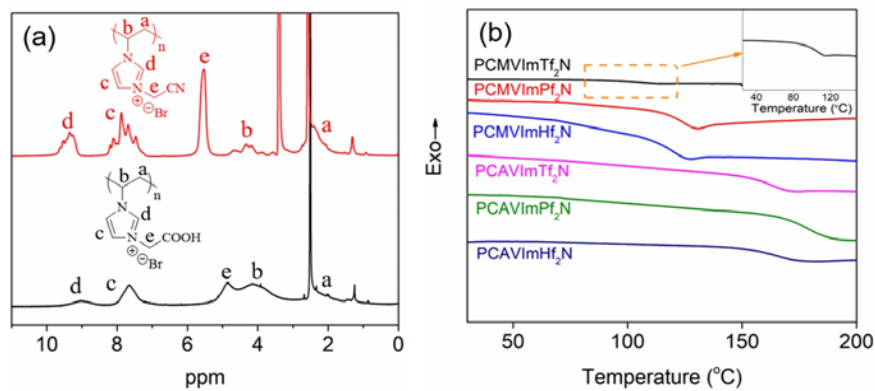

**Figure 2**. (a) $^1$H-NMR spectra of PCMVImBr and PCAVImBr in DMSO-d$_6$. (b) Differential scanning calorimetry traces of PCMVImX and PCAVImX, where x = $Tf_2N$, $Pf_2N$ and $Hf_2N$.

To fabricate all-PIL-based CPMs, we used the PCMVImTf$_2$N/PCAVImTf$_2$N pair as an example. First, a mixture of PCMVImTf$_2$N and PCAVImTf$_2$N was fully dissolved in dimethyl sulfoxide in a 1:1 equivalent molar ratio based on monomer units. The resultant clear solution was then cast onto a glass plate, dried at 80 °C and finally immersed in a 0.05 M aqueous NH$_3$ solution for 4 h to build up a porous membrane, a procedure that is similar to our previous reports.[13] The thin film on the glass was easily peeled off. Analogously, the CPMs composed of PCMVImPf$_2$N/PCAVImPf$_2$N and PCMVImHf$_2$N/PCAVImHf$_2$N were successfully fabricated under the same process. For the sake of clarity, they are termed CPM-x, where x = $Tf_2N$, $Pf_2N$ and $Hf_2N$.

**Figure 3a** is a representative digital photograph of CPM-Tf$_2$N. Note that these CPMs can be readily scaled to fabricate, and the size and thickness were dependent on the amount of PILs used. The function of *aq*. NH$_3$ in this process is to neutralize the COOH unit in PCAVImX, which introduced interpolyelectrolyte complexation between PCMVImX and neutralized

PCAVImX, which is confirmed by the Fourier transform IR spectroscopy **(Figure S12)**. Simultaneously, the diffusion of water molecules into the film induced phase separation of the PCMVImX/PCAVImX blend, creating 3D interconnect porous architectures.[14] Based on this analysis, the chemical structures of CPM-1, CPM-2 and CPM-3 were proposed and shown in **Figure 3b and Figure S13-14.**

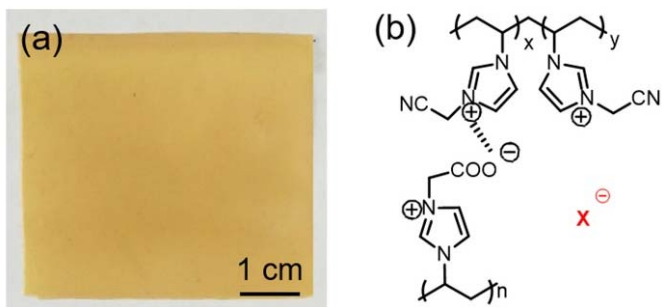

**Figure 3**. (a), (b) A representative digital photograph and the chemical structure of CPM-Tf$_2$N. The ones for CPM-Pf$_2$N and CPM-Hf$_2$N can be found in Figure S13-14.

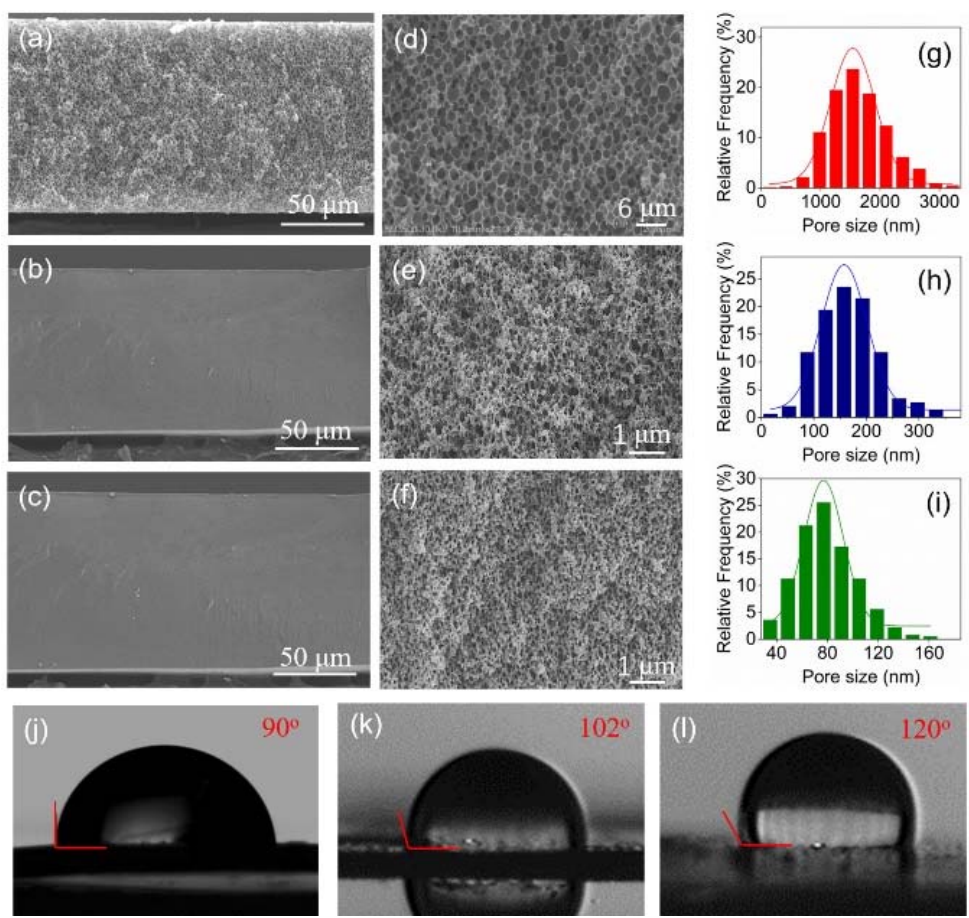

**Figure 4**. (a)-(c), low-magnification cross-sectional SEM images of CPM-1, CPM-2 and CPM-3, respectively. (d)-(f), high-magnification cross-sectional SEM images of CPM-1, CPM-2 and CPM-3, respectively. (g)-(i), water contact angle measurements of CPM-1, CPM-2 and CPM-3, respectively.

- 4 -

**Figures 4a-c** show the low-magnification cross-sectional scanning electron microscopy (SEM) images of CPMs. It can be clearly seen that a uniform porous system was successfully created in CPM-Tf$_2$N (**Figure 4a**), while in **Figures 4b-c**, CPM-Pf$_2$N and CPM-Hf$_2$N appear rather smooth, as their pores are so small thus invisible at this magnification. In their high-magnification SEM images (**Figure 4d-e**), porous architectures were identified in all CPMs. The average pore size gradually decreased from CPM-Tf$_2$N (1546 nm) to CPM-Pf$_2$N (157 nm) and CPM-Hf$_2$N (77 nm) (**Figure 4g-i**). The pore size on the surface of these membranes are similar to their bulk (**Figure S15**). As discussed above, the *aq.* NH$_3$ activation step introduced interpolyelectrolyte complexation between PCMVImX and neutralized PCAVImX, and the water diffusion step induced phase separation of hydrophobic PCMVImX, both of which are responsible for the creation of pores. This process was apparently governed by the diffusion rate of aq. NH$_3$ solution into the films and the concurrent phase separation scale. Extending fluorinated alkyl chains from Tf$_2$N to Pf$_2$N and Hf$_2$N, the hydrophobicity of PIL-X increases in the order of PILTf$_2$N < PILPf$_2$N < PILHf$_2$N. With increasing hydrophobicity of CPM-Tf$_2$N to CPM-Pf$_2$N and CPM-Hf$_2$N, the diffusion rate of aq. NH$_3$ solution was more retarded, which slowed down the water diffusion rate so the interpolyelectrolyte complexation process driven by NH$_3$ diffusion was dominant and locked up at their smaller pore size scale.

The water contact angles of CPM-Tf$_2$N, CPM-Pf$_2$N and CPM-Hf$_2$N were measured to be 90°, 102° and 120°, respectively. It means the wettability of CPM surface can be readily tuned by rational choice of counter-anions. Together, utilizing different anions of PILs endows the synergy between phase separation and aq. NH$_3$ diffusion rate, which ultimately leads to a tailorable porous architecture and surface wettability.

Functional membranes with designable wettability (hydrophobic/hydrophilic) have huge potential in practical applications including separation,[15] antifogging,[16] self-cleaning,[17] anticorrosion,[18] water harvesting biomolecular interactions[19] and microfluidics[20] *etc*. Generally, the degree of hydrophilicity/hydrophobicity is determined merely by the membrane surface chemistry, which is associated with the surface chemical compositions and morphologies (such as pore size).[21] Various techniques have been previously attempted to construct membranes of specific porous architecture and wettability, including chemical vapor deposition,[22] corona discharge,[23] gradual immersion,[24] layer-by-layer deposition technique,[25] hydrothermal route[26] and so on. These techniques are difficult to simultaneously tune the pore size in membrane bulk and membrane surface wettability. In this work, we provide a novel all-PIL materials solution to control both in a straightforward manner.

## CONCLUSIONS

In summary, we have developed an all-PIL based CPM fabrication method. The pore sizes and surface wettability of CPMs could be simultaneously tuned by choice of the counter-anions of PILs with different hydrophobicity. In view of the chemical diversity of PILs, it is believed that a wide range of useful CPMs with desirable charge feature, wettability and pore architecture could be accessed to address, especially current energy and environmental issues.

## SUPPORTING INFORMATION

Supporting Information is available from the Wiley Online Library or from the author

## ACKNOWLEDGEMENT

H. W. acknowledges the financial support from the Nankai University and National Science Foundation of China (No. 21875119). J.Y. is grateful for financial support from the ERC Starting Grant NAPOLI -639720 and the Wallenberg Academy Fellow program (KAW 2017.0166). The authors thank Prof. Tianying Guo for the water contact angle measurements.

# Supporting Information for

## 1. Materials

1-Vinylimidazole (Aldrich 99%), azodiisobutyronitrile (AIBN, Aldrich 98%), bromoacetonitrile (Aldrich 97%), lithium bis(trifluoromethane sulfonyl)imide (Aldrich 99%), lithium bis(pentafluoroethyl sulfonyl)imide(LiPf2N, TCI 98%), Lithium bis(heptafluoropropane sulfonyl)imide (LiHf2N, Wako Pure Chemical Industries, Ltd 98%) and bromoacetic acid (Aldrich 97%) were used as received without further purifications. Dimethyl sulfoxide (DMSO), tetrahydrofuran (THF) and ethanol were of analytic grade.

## 2. Characterization and Measurements

The data for NMR spectra were recorded at 293 K on a Bruker AVANCE AV 400 (400MHz), and chemical shifts were recorded relative to the solvent resonance. The morphologies of different poly(ionic liquid) membranes were viewed by field-emission scanning electron microscopy (SEM, JEOL JSM7500F). Water contact angles (WCA) were measured using the sessile drop method on Dataphysics OCA ISEC contact angle meter (German). Thermogravimetric analyses (TGA) were measured on a Perkin Elmer TG/TGA 6300 at a heating rate of 10 °C min$^{-1}$. Differential scanning calorimetry (DSC) measurements were recorded on a DSC 822e thermal analysis system (Mettler Toledo Instruments Inc. Switzerland) at a heating rate of 5 °C /min with nitrogen protected. FT-IR spectra were collected on a Bruker TENSOR II apparatus in the wavenumber region from 4000 to 400 cm$^{-1}$.

## 3. Materials synthesis and their structural characterization

### 3.1 Preparation of 1-cyanomethyl-3-vinylimidazolium bromide monomer

In a 100 ml flask, 1-vinylimidazole (10.0 g, 0.106 mol) and bromoacetonitrile (12.6 g, 0.106 mol) were added into 70.0 ml of acetone. After the mixture was stirred for 24 hours at room temperature, the precipitate was filtered off and washed with diethyl ether, and finally dried under vacuum at room temperature.

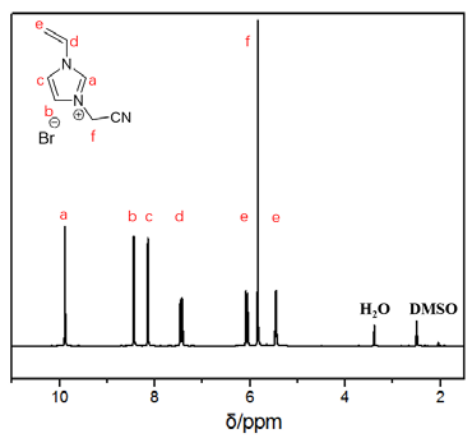

**Figure S1.** $^1$H-NMR spectrum of 1-cyanomethyl-3-vinylimidazolium bromide

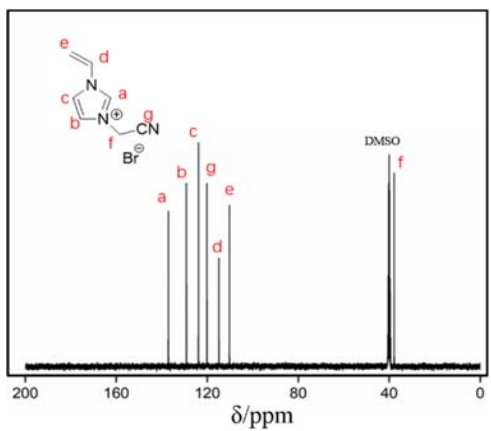

**Figure S2.** $^{13}$C{H}-NMR spectrum of 1-cyanomethyl-3-vinylimidazolium bromide

**3.2 Preparation of 1-carboxymethyl-3-vinylimidazolium bromide monomer**

In a 100 ml flask, 1-vinylimidazole (10.0 g, 0.106 mol) and bromoacetic acid (14.7 g, 0.106 mol) were added into 70.0 ml of acetone. After the mixture was stirred for 24 hours at room temperature, the precipitate was filtered off and washed with diethyl ether and finally dried under vacuum at room temperature. Yield (22g, 92%).

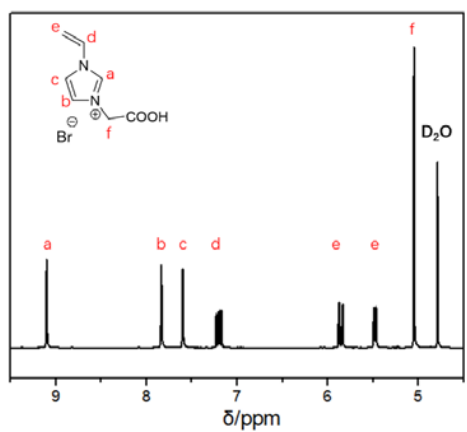

**Figure S3.** $^1$H-NMR spectrum of 1-carboxymethyl-3-vinylimidazolium bromide

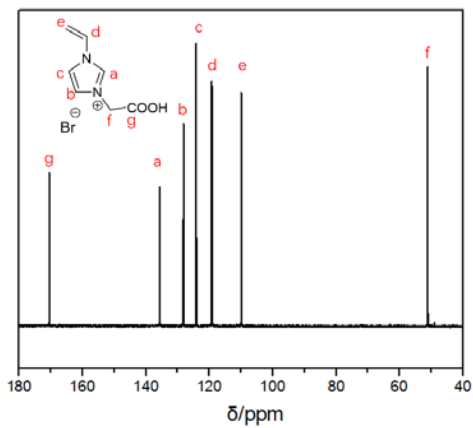

**Figure S4.** $^{13}$C{H}-NMR spectrum of 1-carboxymethyl-3-vinylimidazolium bromide

**3.3 Preparation of poly(1-cyanomethyl-3-vinylimidazolium bromide), PCMVImBr**

10 g of 1-cyanomethyl-3-vinylimidazolium bromide monomer, 0.2 g of AIBN and 100 ml of DMSO were loaded into a 250 ml flask. The mixture was deoxygenated three times by a freeze-pump-thaw procedure and finally charged with nitrogen. The reaction mixture was then placed in an oil bath at 80 °C for 24 hours. When cooling down to room temperature, the reaction mixture was dropwise added to an excess of THF. The precipitate was filtered off, washed with excess of methanol and dried at 90 °C under vacuum.

### 3.4 Preparation of poly(1-carboxymethyl-3-vinylimidazolium bromide), PCAVImBr

10 g of 1-carboxymethyl-3-vinylimidazolium bromide monomer, 0.2 g of AIBN and 100 ml of DMSO were loaded into a 250 ml flask. The mixture was deoxygenated three times by a freeze-pump-thaw procedure and finally charged with nitrogen. The reaction mixture was then placed in an oil bath at 80 °C for 24 hours. When cooling down to room temperature, the reaction mixture was dropwise added to an excess of THF. The precipitate was filtered off, washed with excess of ethanol and dried at 90 °C under vacuum.

### 3.5 Preparation of poly(1-cyanomethyl-3-vinylimidazolium bis(trifluoromethane sulfonyl)imide), PCMVImTf$_2$N

10 g of PCMVImBr was dissolved in 200 ml of deionized water. A 100 ml of aqueous solution of 15 g bis(trifluoromethane sulfonyl)imide lithium salt was added dropwise into the aqueous PCMVImBr solution. After addition, the reaction mixture was allowed to stir for 2 hours and the precipitate was collected by filtration, washed several times with deionized water and dried at 90 °C under vacuum.

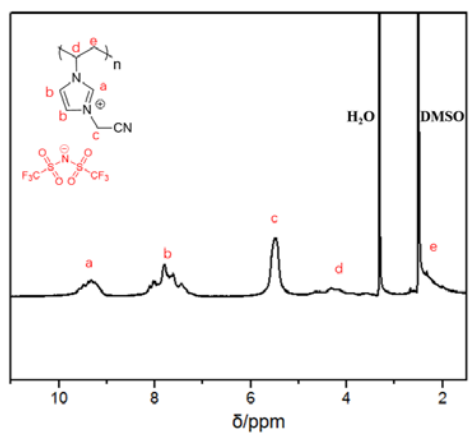

**Figure S5.** $^1$H-NMR spectrum of PCMVImTf$_2$N (DMSO-$d_6$).

### 3.6 Preparation of poly(1-cyanomethyl-3-vinylimidazolium bis(pentafluoroethane sulfonyl)imide), PCMVImPf$_2$N

2 g of PCMVImBr was dissolved in 40 ml of deionized water. A 40 ml of aqueous solution of 4 g bis(pentafluoroethane sulfonyl)imide lithium salt was added dropwise into the aqueous PCMVImBr solution. After addition, the reaction mixture was allowed to stir for 2 hours and the precipitate was collected by filtration, washed several times with deionized water and dried at 90 °C under vacuum.

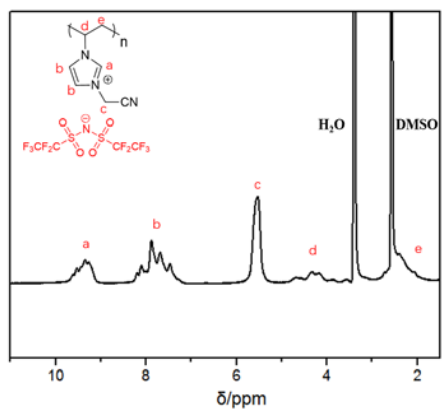

**Figure S6.** $^1$H-NMR spectrum of PCMVImPf$_2$N (DMSO-$d_6$).

## 3.7 Preparation of poly(1-cyanomethyl-3-vinylimidazolium bis(heptafluoropropane sulfonyl)imide), PCMVImHf$_2$N

2 g of PCMVImBr was dissolved in 40 ml of deionized water. A 50 ml of aqueous solution of 5 g bis(heptafluoropropane sulfonyl)imide lithium salt was added dropwise into the aqueous PCMVImBr solution. After addition, the reaction mixture was allowed to stir for 2 hours and the precipitate was collected by filtration, washed several times with deionized water and dried at 90 °C under vacuum.

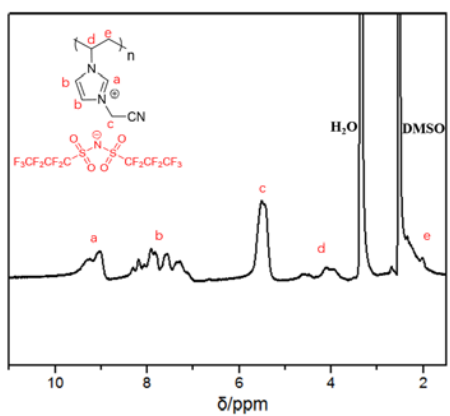

**Figure S7.** $^1$H-NMR spectrum of PCMVImHf$_2$N (DMSO-$d_6$).

## 3.8 Preparation of Poly(1-carboxymethyl-3-vinylimidazolium) bis(trifluoromethane sulfonyl)imide, PCAVImTf$_2$N

10 g of PCAVImBr was dissolved in 200 ml of deionized water. A 100 ml of aqueous solution of 13 g bis(trifluoromethane sulfonyl)imide lithium salt was added dropwise into the aqueous PCAVImBr solution. After addition, the reaction mixture was allowed to stir for 2 hours and the precipitate was collected by filtration, washed several times with deionized water and dried at 90 °C under vacuum.

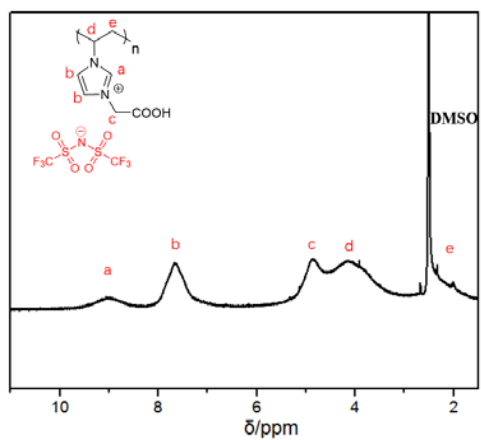

**Figure S8.** $^1$H-NMR spectrum of PCAVImTf$_2$N (DMSO-d$_6$).

### 3.9 Preparation of poly(1-carboxymethyl-3-vinylimidazolium bis(pentafluoroethane sulfonyl)imide), PCAVImPf$_2$N

 2 g of PCAVImBr was dissolved in 40 ml of deionized water. A 40 ml of aqueous solution of 3.6 g bis(pentafluoroethane sulfonyl)imide lithium salt was added dropwise into the aqueous PCAVImBr solution. After addition, the reaction mixture was allowed to stir for 2 hours and the precipitate was collected by filtration, washed several times with deionized water and dried at 90 °C under vacuum.

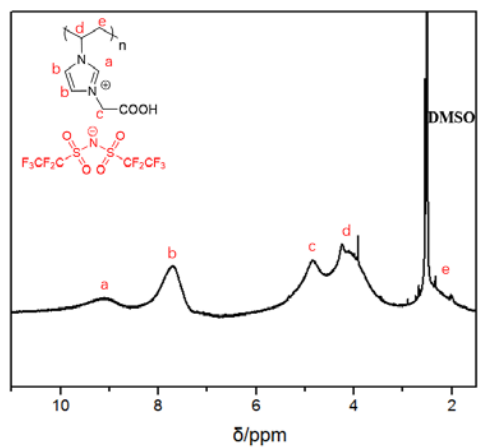

**Figure S9.** $^1$H-NMR spectrum of PCAVImPf$_2$N (DMSO-*d$_6$*).

### 3.10 Preparation of poly(1-carboxymethyl-3-vinylimidazolium bis(heptafluoropropane sulfonyl)imide), PCAVImHf$_2$N

 2 g of PCAVImBr was dissolved in 40 ml of deionized water. A 50 ml of aqueous solution of 4.6 g bis(heptafluoropropane sulfonyl)imide lithium salt was added dropwise into the aqueous PCAVImBr solution. After addition, the reaction mixture was allowed to stir for 2 hours and the precipitate was collected by filtration, washed several times with deionized water and dried at 90 °C under vacuum.

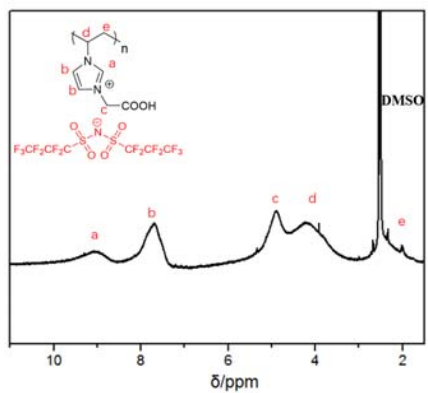

**Figure S10.** $^1$H-NMR spectrum of PCAVImHf$_2$N (DMSO-$d_6$).

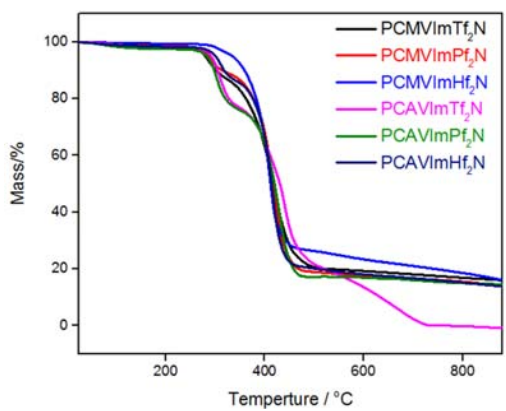

**Figure S11.** TGA traces of PCMImX and PCAImX, where X=Tf$_2$N, Pf$_2$N and Hf$_2$N respectively.

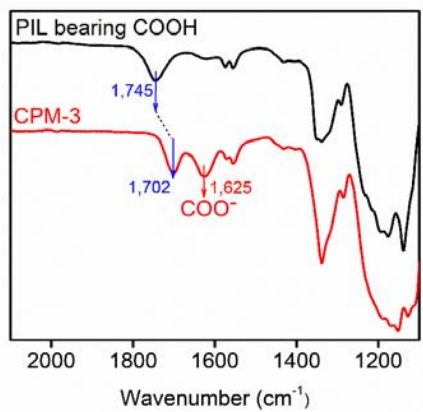

**Figure S12.** Representative FT-IR spectra of PIL-bearing COOH and CPM-3. It can be clearly seen that a new deprotonated COO-signal (C=O) at 1625 cm$^{-1}$ can be detected in CPM-3. Furthermore, due to the delocalization of COO- moiety, the original (C=O) signal at 1745 cm$^{-1}$ shift to 1702 cm$^{-1}$. These results obviously demosntrated the presence of deprotonated COOH in CPM-3 by aq NH$_3$ solution treatment.

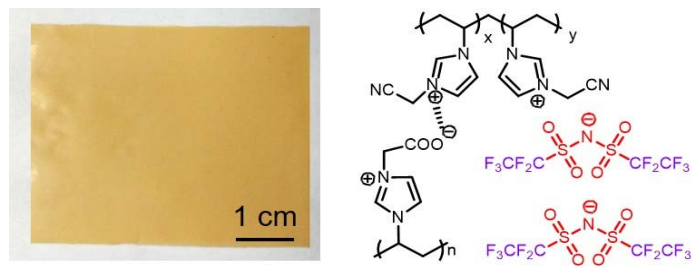

**Figure S13.** Digital photograph and the chemical structure of of CPM-Pf₂N.

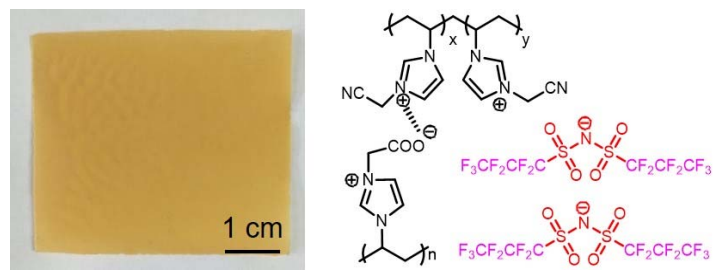

**Figure S14.** Digital photograph and the chemical structure of of CPM-Hf₂N.

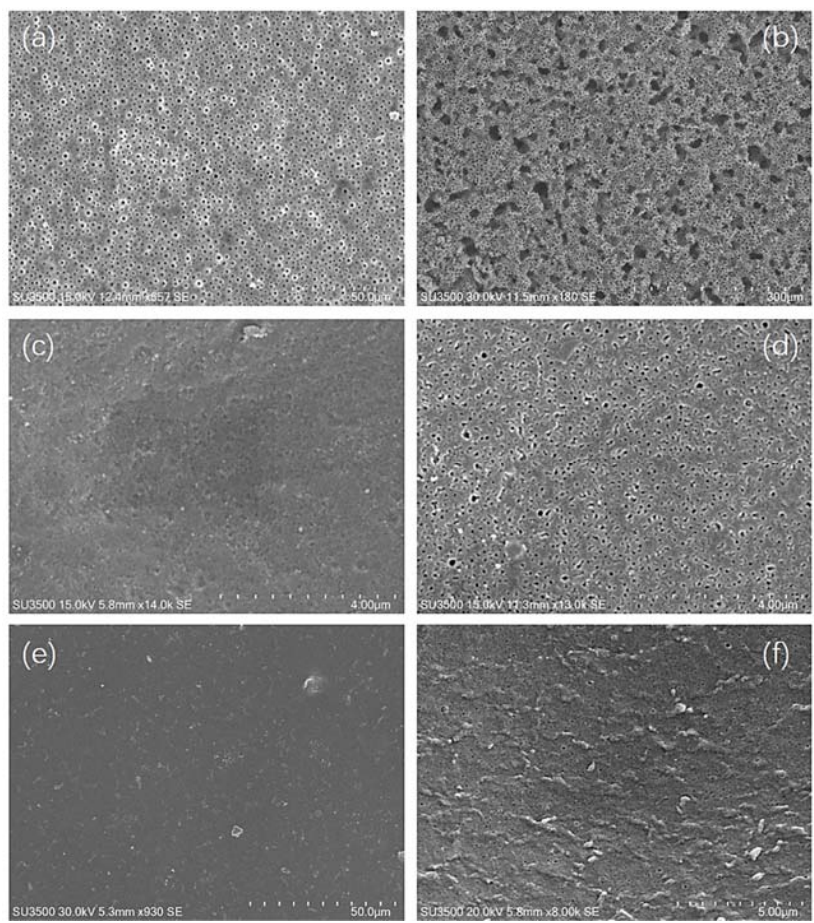

**Figure S15.** SEM images of the top surface of (a) CPM-Tf₂N, (c) CPM-Pf₂N and (e) CPM-Hf₂N, respectively, and the bottom surface of (b) CPM-Tf₂N, (c) CPM-Pf₂N and (f) CPM-Hf₂N, respectively. Note: the membrane surfaces facing the air and the substrate are named the top and bottom surfaces, respectively.